\documentstyle[manuscript,eqsecnum,aps,version2]{revtex}
\draft

\begin{document}

\title
\bf Finite Temperature Quark Confinement
\endtitle
\author{Peter~N.~Meisinger and Michael~C.~Ogilvie}
\instit
Department of Physics, Washington University, St. Louis, MO 63130
\endinstit
\medskip
\centerline{\today}

\abstract

Confinement may be more easily demonstrated at finite temperature
using the Polyakov loop
than at zero temperature using the Wilson loop.
A natural mechanism for confinement can arise via the coupling
of the adjoint Polyakov loop to $ F_{\mu\nu}^2$.
We demonstrate this mechanism
with a one-loop calculation of
the effective potential for $SU(2)$ gluons in a background field
consisting of a non-zero color magnetic field and a non-trivial
Polyakov loop. The color magnetic field drives the Polyakov loop
to non-trivial behavior, and the Polyakov loop can remove the
well-known tachyonic mode associated with the Saviddy vacuum.
Minimizing the real part of the effective potential leads to
confinement, as determined by the Polyakov loop. 
Unfortunately, we cannot arrange for simultaneous stability and confinement
for this simple class of field configurations.
We show for a large class of 
abelian background fields
that at one loop tachyonic modes are necessary for confinement.

\endabstract

\pacs{PACS numbers: 11.10.Wx 11.30.Qc 12.38.Aw}

\pagebreak[4]

\section{INTRODUCTION}
\label{s1}

While the problem of quark confinement has been with us for some time
\cite{Mandelstam}, a completely satisfactory explanation of the
mechanism has been elusive.
It is widely held
that unbroken non-Abelian gauge theories functions as dual superconductors,
with some analog of the Meissner effect in type II superconductors
confining chromoelectric flux into flux tubes.
A fully satisfactory theory of quark confinement
would be Lorentz invariant and gauge invariant, and have
correct renormalization group behavior, which suggests that
a successful theory must maintain a tight connection with
perturbation theory.

Confinement is easiest to
demonstrate at finite temperature, where it
can be studied using the trace of the Polyakov loop
$ {\cal P} = {\cal T} \exp \left[ i \int_0^\beta dt A_0 \right] $.
In pure gauge theories, without fundamental representation quarks,
confinement is equivalent to 
the vanishing of the expected value of the trace of the Polyakov loop
in the fundamental representation,
$ Tr_F {\cal P} = 0$.
In contrast, demonstrating confinement at zero temperature
requires showing that the Wilson loop has area law behavior
for large areas. To use a simple analogy, it is much easier
to study one-point functions than two-point functions.
In pure $SU(N)$ gauge theories, there is a global $Z(N)$ symmetry,
which, if unbroken, enforces the vanishing of the Polyakov loop.
The spontaneous breaking of this $Z(N)$ symmetry at high temperature
is observed in lattice gauge theory simulations. 
In accord with these results, a successful
theory of confinement should also show that
the expected value of the trace of the Polyakov loop
vanishes at low temperatures, and is non-zero at
high temperatures.

The perturbative effective potential for the Polyakov loop,
or equivalently, $A_0$ in an appropriate gauge
does not indicate confinement \cite{WeI,WeII}.
The general result for $SU(N)$ is a quartic polynomial in
$A_0$.
The minimum of this effective potential is always at $A_0 = 0$,
corresponding to $tr_F({\cal P}) = N$; in the pure gauge theory
without quarks, there are also equivalent solutions related by
the global $Z(N)$ symmetry.
There is a simple physical reason behind this: when $A_0 = 0$,
the free energy associated with the quark-gluon plasma is minimized.

One possible origin of confinement is that at low temperatures
different regions of space are in different
$Z_N$ phases, in such a way that $tr_F({\cal P})$
averages to zero.
This possibility is realized in the strong-coupling region of $Z(N)$ lattice
gauge theories, where individual Polyakov loops can only take on values
in $Z(N)$, but the average over an infinite volume is zero. 
This $Z(N)$ confining phase is connected to a phase of $SU(N)$ lattice
gauge theories: if the Wilson action $\beta_F S_F$ has added to it
an adjoint term $\beta_A S_A$, then the $Z(N)$ gauge theory can be obtained
in the limit $\beta_A \to \infty$ with $\beta_F$ fixed. The region
$\beta_A$ large and $\beta_F$ small realizes confinement in the same manner
as the $Z(N)$ model. However, this region of the $\beta_F$-$\beta_A$ plane
is separated from the continuum limit fixed point by a line of first-order
phase transitions\cite{BhanotCreutz}.
In contrast, examination of lattice field configurations
generated by Monte Carlo simulation 
along the $\beta_A = 0$ axis indicate that
the magnitude of individual Polyakov loops 
is small at low temperatures and increases
dramatically at the confinement transition.
This argues against the possibility that
confinement occurs because different regions of space are in different
$Z_N$ phases.

We advocate another possibility: that the coupling between the local
gauge field $F_{\mu\nu}$ and the adjoint Polyakov loop produces an effective
action which leads to two different phases. In the
low temperature phase, there is a magnetic condensate and the Polyakov loop
indicates confinement; in the high temperature phase the magnetic condensate
vanishes, the Polyakov loop indicates deconfinement, and
the contribution of the thermal gauge boson gas dominates
the free energy. 
This occurs because finite temperature effects from gluons
naturally couple the local gauge field to adjoint representation
Polyakov loops\cite{MeOgI,MeOgII,MeOgIII,MeOgIV}.

In order to produce confinement, the low temperature phase
must minimize its free energy by minimizing the expected value
of the trace of the adjoint Polyakov loop.
For $SU(N)$,
the fundamental and adjoint representation Polyakov loops are related by
\begin{eqnarray}
Tr_A ( {\cal P} ) = | Tr_F ( {\cal P} ) |^2 - 1.
\end{eqnarray}
Clearly, when $Tr_A ( {\cal P} )$ assumes its minimum value of $-1$,
$Tr_F ( {\cal P} )$ assumes the value $0$. 
As seen in previous calculations, 
at sufficiently high temperature, contributions from thermal gluons will
dominate the
free energy, driving the Polyakov loop towards its maximum value.
In our scenario, this also drives
the magnetic condensate to zero.
The coupling of the gauge field to the adjoint Polyakov loop
naturally 
incorporates the behavior known from 
lattice simulations: both the Polyakov loop and
plaquette expectation values change at the deconfining phase transition
of $SU(N)$ lattice gauge theories.

The idea that more than one field must be considered to describe
the behavior of finite temperature QCD has previously been 
proposed by
Campbell, Ellis and Olive \cite{CaElOlI,CaElOlII}.
In their effective theory, a phenomenological
coupling of the gluon condensate 
to the chiral condensate was postulated.
However, this approch does not include the important role
played by the Polyalov loop in finite temperature chiral behavior.
It is easy to show that there is a 
coupling between the chiral condensate and Polyakov loops.
For example in references \cite{MeOgI} and \cite{MeOgII},
we showed using lattice perturbation theory
the role that the Polyakov loop
plays in finite temperature corrections to the renormalization of
the gauge field coupling constant. Differentiation of our result with
respect to the quark mass shows perturbatively that the finite
temperature corrections to the chiral condensate
depend on the Polyakov loop as well as the gluon condensate.
We have also shown in detail 
the interplay between the Polyakov loop and the chiral condensate 
for a variant of the Nambu-Jona-Lasinio model.
\cite{MeOgIII,MeOgIV}.
A truly satisfactory treatment of the critical behavior of finite
temperature QCD, including the effects of quarks,
is likely to require treatment of the gluon condensate,
the chiral condensate and the Polyakov loop.

\section{CALCULATION OF THE FREE ENERGY}
\label{s2}

The observation of Saviddy \cite{Saviddy} that the perturbative QCD vacuum
is unstable with respect to the formation of a constant chromomagnetic
field is one proposed origin for a magnetic condensate in QCD.
However, Nielsen and Olesen \cite{NiOl} showed that such field
configurations are themselves unstable, due to a tachyonic mode
in the one-loop determinant. Several authors 
\cite{NiSa,Muller,Dittrich,Kapusta}
have extended the work
of Nielsen and Olesen to finite temperature, treating the tachonic
modes in various ways, including simply ignoring them.

We demonstrate the main points of
our proposed mechanism by incorporating the effects of 
a non-trivival Polyakov loop, treating all modes exactly at one loop.
The tachyonic mode can be removed by a non-trivial
Polyakov loop.
The free energy of an SU(2) gluon gas moving in a background
field consisting of a constant color magnetic field and and a constant
non-trivial Polyakov loop is obtained by calculating the one-loop,
finite temperature effective
potential in the background field gauge; see \cite{Ab} and \cite{PeSc}
for pedagogical introductions to the background field method.
This is equivalent
to summing up three contributions to the free energy: 
the classical action, the zero-point energies, and the
free energy of the gluon gas. All finite temperature effects
reside in the last term.
As we have shown elsewhere, Polyakov loops appear naturally
at one loop in an image expansion of finite temperature determinants
\cite{MeOgI,MeOgII,MeOgIII,MeOgIV}. 
Here, similar results are achieved by directly computing the functional
determinant.

We choose the color magnetic field
and the Polyakov loop to be simultaneously diagonal, and we
let the color magnetic field $H$ point in the $x_3$ direction.
We take the Polyakov loop to be specified by a constant $A_0$ field,
given in the adjoint representation by
\begin{eqnarray}
A_0 = {\phi \over {2 \beta}} \tau_3.
\label{e2.a}
\end{eqnarray}
The trace of the Polyakov loop is then given by
\begin{eqnarray}
Tr_F ( {\cal P} ) = 2 \cos (\phi / 2 )
\label{e2.b}
\end{eqnarray}
in the fundamental representation and by
\begin{eqnarray}
Tr_A ( {\cal P} ) = 1 + 2 \cos (\phi )
\label{e2.c}
\end{eqnarray}
in the adjoint representation.
The external magnetic field we take to have the form
\begin{eqnarray}
A_2 = {1 \over 2} H x_1 \tau_3
\label{foo}
\end{eqnarray}
which gives rise to a chromomagnetic field
\begin{eqnarray}
F_{12} = {1 \over 2} H \tau_3
\label{bar}
\end{eqnarray}
As explained, for example, in \cite{NiOl}, the external field
gives rise to Landau levels in the gluon functional determinant.

The one-loop contribution to the free energy 
has the form \cite{NiOl,NiSa}:
\begin{eqnarray}
&\phantom{+}& 2 \times {1 \over 2} \, \sum_{m = 0}^{\infty} \,
\sum_{n, \pm} \, {1 \over \beta} \,
\, { H \over 2 \pi} \, \int \, {dk_3 \over 2 \pi } \, 
\ln \left[ \left( \omega_n - {\phi \over \beta} \right)^2 +
2 H \left( m + {1 \over 2} \pm 1 \right)  + k_3^2 \right]
\nonumber\\
&+&  \sum_n \, {1 \over \beta} \,
\int \, {d^3 \vec{k} \over ( 2 \pi )^3 } \, 
\ln \left( {\omega}_n^2 + \vec{k}^2 \right),
\label{e2.1}
\end{eqnarray}
where the $\omega_n = 2 \pi n / \beta$ are the usual Matsubara frequencies, and
where the terms $2 H (m + 1/2 \pm 1)$ are the allowed Landau levels of the
gauge field with its spin coupled to H. The explicit factor of $2$ in front of
the first term in the sum results from the trace over color degrees of
freedom.

When $ \phi = 0$,
the $n = 0$ and $m = 0$ modes give rise to tachyonic modes
for $k_3$ sufficiently small;
these in turn give rise to an imaginary part in the free energy.
We observe that these same modes will give a strictly real
factor to the determinant provided
\begin{eqnarray}
\phi \geq \beta \sqrt{H}.
\label{e2.2}
\end{eqnarray}
However, if $\phi$ is too large, it may cause the $n = 1$ mode to become
unstable. Hence, the general criterion for stability is
\begin{eqnarray}
\beta \sqrt{H} < \phi < 2 \pi -  \beta \sqrt{H}
\label{e2.3}
\end{eqnarray}.

A standard product representation \cite{GrRy}
\begin{eqnarray}
\prod_k \left[ 1 + \left( {x \over 2 k \pi - a} \right)^2 \right] =
{cosh(x) - cos(a) \over 1 - cos(a)}
\label{e2.4}
\end{eqnarray}
allows us to write the effective potential in the form:
\begin{eqnarray}
V = {1 \over 2 g^2 } H^2 & + &
{1 \over 2} \, \sum_{m = 0}^{\infty} \, \sum_c \,
\, { H \over 2 \pi \beta} \, \int \, {dk \over 2 \pi } \, 
\ln \left\{ \cosh \left[ \beta {\omega}_{\pm} (m, k) \right] -
\cos (\beta \phi ) \right\}
\nonumber\\
&+& {1 \over \beta} \,
\int \, {d^3 \vec{k} \over ( 2 \pi )^3 } \, 
\ln \left\{ \cosh \left[ \beta |\vec{k}| \right] - 1\right\} ,
\label{e2.7}
\end{eqnarray}
where
\begin{eqnarray}
{\omega}_{\pm}^2 (m, k) = 
2 H \left( m + {1 \over 2} \pm 1 \right)  + k^2
\label{e2.8}
\end{eqnarray}
which can be negative.
Note that the classical free energy of the magnetic field, $ H^2 / 2g^2$, is
included in this expression. Expanding the logarithm and discarding an
irrelevant constant, 
the contribution of the Landau levels to the 
the effective potential can also be written in the form:
\begin{eqnarray}
{1 \over 2} \, \sum_{m = 0}^{\infty} \, \sum_c \,
\, {H \over 2 \pi} \, \int \, {dk \over 2 \pi } \, 
\left[ {\omega}_{\pm} (m, k) - {2 \over \beta} \, \sum_{n = 1}^{\infty} \,
{\cos (n {\phi}^c) \over n} e^{- n \beta
{\omega}_{\pm} (m, k)} \right].
\label{e2.9}
\end{eqnarray}

After performing the sum over $m$ in the equation above, we
regulate the divergent portions of the resulting integral
using the identity \cite{NiOl}
\begin{eqnarray}
( {\omega}^2 - i \delta )^{\nu - \mu} =
{i^{\mu - \nu} \over \Gamma(\mu - \nu)} \, \int \, d\tau \,
{\tau}^{\mu - \nu - 1} \, e^{- i \tau ({\omega}^2 - i \delta)}.
\label{e2.10}
\end{eqnarray}
This yields a renormalized effective potential with real component
\begin{eqnarray}
V_R &=& {11  H^2 \over 48 {\pi}^2 } \,
\ln \left( {H \over {\mu}_0^2} \right) -
{ ( H )^{3/2} \over {\pi}^2 \beta} \, \sum_{n = 1}^{\infty} \,
{\cos (n \phi) \over n} \left[ K_1 (n \beta \sqrt{H} ) - {\pi \over 2}
Y_1 (n \beta \sqrt{H} ) \right]
\nonumber\\
&\phantom{=}& - {2 ( H )^{3/2} \over {\pi}^2 \beta} \, \sum_{n = 1}^{\infty}
\, {\cos (n \phi) \over n} \, \sum_{m = 0}^{\infty} \,
\sqrt{2 m + 3} K_1 [n \beta \sqrt{(2 m + 3)H} ]
- { 2 \pi^2 \over 90 \beta^4 }
\label{e2.11}
\end{eqnarray}
and imaginary component
\begin{eqnarray}
V_I = - { H^2 \over 8 \pi} - {H^{3/2} \over 2 \pi \beta} \,
\sum_{n = 1}^{\infty} \, {\cos (n \phi ) \over n} \,
J_1 ( n \beta \sqrt{H} ),
\label{e2.12}
\end{eqnarray}
where $\phi$ is defined in Eq.~(\ref{e2.a}).

It is far from obvious that the potential we have found is
real when the constraint of Eq.~(\ref{e2.3}) is satified. We have
checked numerically that the imaginary part of the potential
behaves as it should. In figure 1, we show the imaginary part
for $\sqrt{H}/\mu = 0.5$ and $T/\mu = 0.25$. The theory is stable in the
region given by Eq.~(\ref{e2.3}).

A similar
calculation has been performed for different reasons by
Starinets, Vshivtsev, and Zhukovskii \cite{StVsZh}.
Our results are in disagreement, with Bessel functions
interchanged between the real and imaginary part.
However, our results are in numerical agreement with the
exact result for $V_I$ derived by 
Cabo, Kalashnikov, and Shabad
\cite{CaKaSh}
for the case $\phi=0$,
and our result for $V_I$ is zero for the 
range of $\phi$ determined by Eq.~(\ref{e2.3}).
We are confident 
our results are correct.

Note that
we have carefully chosen the overall additive constant of the free energy
such that $V=0$ when $H=0$ and $T=0$.

\section{Minimization of Re(V)}
\label{s4}

We are interested in showing that the system can lower its free
energy by having $\phi=\pi$, or equivalently $Tr_F P = 0$.
The simplest scheme is to minimize $Re(V)$ over all allowed
values of $\phi$ and $H$, ignoring the imaginary part of $V$.
If the global minimum of $Re(V)$ gives rise to an imaginary part,
that will indicate that this field configuration is unstable,
however.

Some insight into this issue can be obtained by noting that
for very low temperatures, the finite temperature contributions to
$Re(V)$ will be dominated by the $Y_1 (n \beta \sqrt{H})$ terms.
Using the asymptotic form
for large argument, we see that they will favor the adjoint Polyakov loop
being at its minimum, and hence favor confinement.

We have examined this issue numerically and found that at low
temperatures, minimization of $Re(V)$ leads to $\phi=\pi$ being
preferred. This is shown in figure 2, which plots the real part of the
effective potential versus $H/\mu^2$ at $T/\mu = 0.25$
for both $\phi=0$ and $\phi=\pi$.
Figure 3 plots $V_I$, the imaginary part of the effective potential
for the same parameters.
Unfortunately,
examination of $V_I$ shows that the lowest minima is not stable,
lying just to the right of the stable region.

The Saviddy vacuum is preferred over the perturbative vacuum
only for sufficiently low temperatures. 
Figure 4 shows the behavior of $V$ at $T/\mu = 0.73$ and $\phi = \pi$
compared with $V_{0T}$, the one-loop effective potential for 
a gluon gas at $H = 0$ and $A_0 = 0$.
Above the pseudocritical temperature at approximately
$ T_c /\mu = 0.73 $, the perturbative solution with $H=0$ and
$\phi=0$ has lower free energy. Naively, this
would indicate a first-order phase transition.
However, the magnitude of $H/\mu^2$ is so large that a one-loop
calculation cannot be considered reliable.
Further, examination of $V_I$ shows that the constant field
configuration which minimizes $V$ is always unstable. All
that can be said is that at low temperatures, the perturbative solution is
definitely not the state of lowest free energy.

It is amusing to note that the behavior at very low temperatures
reveals the existence of a familiar phenomenon associated with
Landau levels, the de Haas-van Alphen effect. Fig. 5 plots the
free energy as a function of $H/\mu^2$ at $T/\mu = 0.1$ for both
$\phi=0$ and $\phi=\pi$. The minima are the signs of the characteristic
oscillatory components of the magnetic susceptibility.

Because of the possibility of fixing the external field H in
lattice simulations, it may be of interest to examine the
analytical behavior at fixed H.
At fixed H, $\phi = \pi$ is not necessarily the minimum
value of $V$. This is demonstrated clearly in
Fig. 6 and Fig. 7, which show the $V$ and $V_I$
as a function of $\phi$ at $T/\mu = 0.25$ for
$ H/\mu = 0.40$,$0.65$ and $0.90$. For 
$H/\mu^2 = 0.40$, there is a region of $\phi$ where
$V_I$ vanishes.
The minimum of $V$ lies at or very near the boundary
where $V_I$ develops an imaginary part.
This particular case appears similar to the behavior
found by van Baal, who examined self-dual solutions 
on a hypercube in the
presence of a non-trivial Polyakov loop \cite{vanBaal}.

\section{GENERALIZATIONS}
\label{s5}

As we have seen above, the terms in the free energy which favor
confinement come from the contribution of the 
would-be tachyonic modes to the free energy.
We can generalize this result slightly. 
Consider an arbitrary time-independent vector potential
of the form
\begin{eqnarray}
\vec{A}(\vec{x}) \tau_3
\end{eqnarray}
We write the eigenvalues that enter into the functional
determinant in the form
\begin{eqnarray}
\left[ \left( \omega_n - {\phi \over \beta)} \right)^2 +
{\omega}^2 \right]
\end{eqnarray}
\label{e5.5}
where $\omega^2$ can be positive or negative.

We write the 
positive eigenvalues as ${\epsilon}^2$, and the
negative eigenvalues as $- {\chi}^2$.
Following the identical reasoning used above, the positive
eigenvalues contribute to the free energy a term

\begin{eqnarray}
V_{+} = \sum_{n} \, {1 \over \beta} \, \int \, d{\epsilon} \, \rho_{+}
(\epsilon)
\, \ln \left[ \left( \omega_n - {\phi \over \beta} \right)^2 +
{\epsilon}^2 \right],
\label{e5.1}
\end{eqnarray}

where $\rho_{+}(\epsilon)$ is the eigenvalue density. It it easy to see
that such eigenvalues always favor $\phi = 0$. On the other hand, the
negative eigenvalues contribute to the free energy a term

\begin{eqnarray}
V_{-} &=& \sum_{n} \, {1 \over \beta} \, \int \, d{\chi} \, \rho (\chi)
\, \ln \left[ \left( \omega_n - {{\phi} \over \beta)} \right)^2 -
{\chi}^2 \right] \nonumber\\
    &=& C + \int \, d{\chi} \, \rho(\chi)
        \, \ln \left[ \cos^2 \left( \beta \chi \right) -
        \cos^2 \left( {\phi} \right) \right],
\label{e5.2}
\end{eqnarray}
and such terms may favor ${\phi} \neq 0$. Thus in this simple generalization
we have an indication of the likely importance of the negative modes
in bringing about confinement.

It is interesting to note that 
the hard thermal loop resummation of Braaten and
Pisarski \cite{BrPi} drops precisely those features which we are advocating
as the origin of confinement at low temperatures.
The tachyonic
modes in the Saviddy vacuum arise at the one-loop level from the
spin coupling of the gauge field to the external magnetic field. However,
the term accounting for this spin coupling is discarded in the hard thermal
loop approximation, since the term is manifestly soft.

\section{CONCLUSIONS}
\label{s6}

We have seen that 
finite temperature
gauge theory is the best place to prove confinement,
because one need only show that the expected value of the
Polyakov loop vanishes. Models and mechanisms for which this cannot be
demonstrated are not good candidate explanations of confinement.
We have demonstrated these considerations
with a one-loop calculation of
the effective potential for $SU(2)$ gluons in a special background field
consisting of a non-zero color magnetic field and a non-trivial
Polyakov loop. The color magnetic field drives the Polyakov loop
to non-trivial behavior, and the Polyakov loop can remove the
well-known tachyonic mode associated with the Saviddy vacuum.
Minimizing the real part of the effective potential leads to
confinement, as determined by the Polyakov loop.
Unfortunately, we cannot arrange for simultaneous stability and confinement
in this simple model. We are able to show 
for a class of abelian field configurations
that tachyonic modes are necessary for confinement.

The mechanism we have described here for pure gauge theories is
also applicable when the effects of quarks are included. The
quark determinant has several effects, the most important of which
is the explicit breaking of $Z(N)$ symmetry.
Pure $SU(N)$ gauge theories have an exact global $Z(N)$ symmetry
which, if unbroken, requires that the expectation value of
$Tr_F P$ vanish. Fundamental representation particles, such as
quarks, explicitly break this symmetry. However, explicit perturbative
calculations \cite{MeOgII}
shows that the symmetry breaking terms are numerically
small, even when the current quark mass is zero.
Strong-coupling arguments \cite{GockschOgilvie}
suggest that in the low temperature phase, it is the constituent
quark mass rather than the current mass which controls the strength
of the $Z(N)$ symmetry breaking, which would further reduce the effect.

Other effects of quarks enter from the change in renormalization group
beta function coefficients, and from the increase in the number of degrees
of freedom in the high temperature phase. 
At low temperatures, we expect that the competition between the gluonic
terms which drive the trace of the Polyakov loop towards zero and
the quark terms which drive it away from zero will produce a low temperature
region where the Polyakov loop is small, in accord with lattice simulations.

It is obvious that the ground state of QCD is not well modeled by
a constant chromomagnetic field. However, we can extend the insight
we have gained into the possible origins of confinement by examining
the confining properties of other background field configurations.
There are several interesting classes of configurations, including
constant field potentials. If the different space time components
do not commute, this leads to constant color fields which are inequivalent
to the class studied here \cite{BrownW}.
Another interesting class of background field
configurations is abelian monopole configurations. It would also be
very interesting to know if a suitably chosen ensemble of random
background field configurations confines.

\section*{ACKNOWLEDGEMENTS}
We wish to thank the U.S. Department of Energy for financial support under
grant number DE-FG02-91-ER40628.

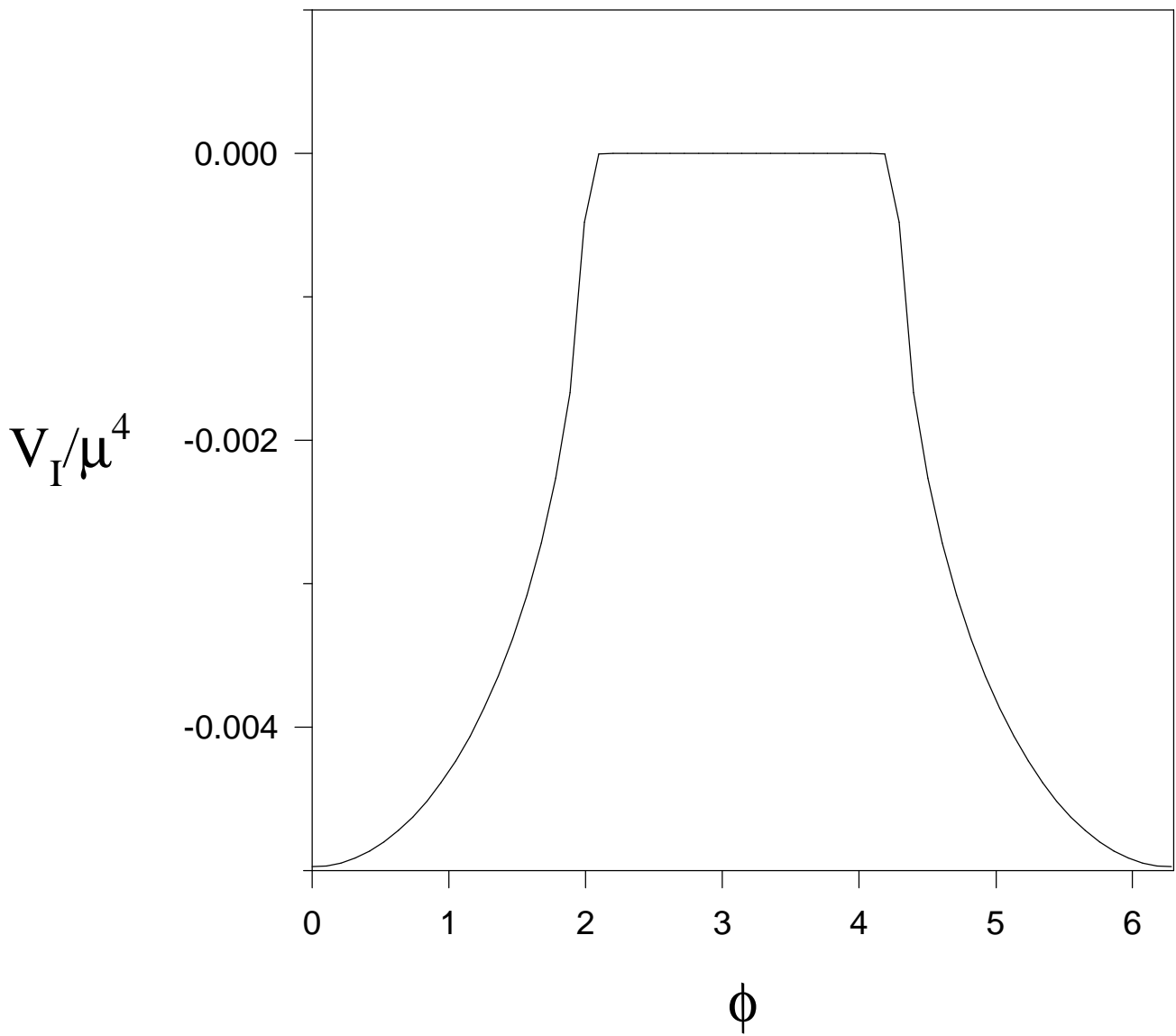
\figure{ $V_I$ as a function of $\phi$ for $H^{1/2}/\mu = 0.5$
and $T/\mu = 0.25$.
\label{f1} }
\figure{ $V$ at $T/\mu = 0.25$ for $\phi = 0 $ and $phi =  \pi$.
\label{f2} }
\figure{ $V_I$ at $T/\mu = 0.25$ for $\phi = 0 $ and $phi =  \pi$.
\label{f3} }
\figure{ $V$ at $T/\mu = 0.1$ for $\phi = 0 $ and $phi =  \pi$.
\label{f4} }
\figure{ $V$ at $T/\mu = 0.73$ for $\phi =  \pi$
compared with $V_{0T}$.
\label{f5} }
\figure{ $V$ as a function of $\phi$ at $T/\mu = 0.25$.
\label{f6} }
\figure{ $V_I$ as a function of $\phi$ at $T/\mu = 0.25$.
\label{f7} }

\end{document}